\documentclass[preprints,accept,oneauthor,pdftex]{Definitions/mdpi} 

\firstpage{1} 
\pubvolume{1}
\issuenum{1}
\articlenumber{0}
\pubyear{2023}
\copyrightyear{2023}
\externaleditor{Academic Editor: Prof. Chengmin Zhang; Guest Editors: Prof. Dr. David Blaschke and Dr. Jerome Petri}
\datereceived{} 
\dateaccepted{} 
\datepublished{} 
\hreflink{https://doi.org/} % If needed use \linebreak
\usepackage{natbib}
\usepackage{color}
\usepackage[english]{babel}
\usepackage{amsmath}
\usepackage{amssymb}
\usepackage{bm}
\usepackage[utf8]{inputenc}
\usepackage{ulem}
\usepackage{cancel}
\usepackage{hyperref}
\usepackage{amsfonts}% or latexsym, amssymb, mathabx, txfonts, pxfonts, wasysym
\newcommand{\gcc}{\mbox{g~cm$^{-3}$}}

\newcommand{\B}{\bm{B}}
\newcommand{\vv}{\bm{v}}
\newcommand{\vu}{\bm{u}}
\newcommand{\vr}{\bm{r}}
\newcommand{\dd}{\mathrm{d}}

\newcommand{\Msun}{M_\odot}

\newcommand{\msun}{{M}_\odot}

\Title{Zeeman Splitting of Torsional Oscillation
	Frequencies of Magnetars}

\TitleCitation{Zeeman splitting of torsional oscillations}
\Author{Dmitry Yakovlev 
	$^{\dagger}$
\orcidA{} 
}

\AuthorNames{Dmitry Yakovlev}

\AuthorCitation{Yakovlev, D.}
\address[1]{%
 Ioffe Institute, Politekhnicheskaya 26, 194021 St~Petersburg,
	Russia; yak@astro.ioffe.ru
}

\corres{Correspondence: yak@astro.ioffe.ru}

\abstract{Magnetars form a special class of neutron stars 
possessing superstrong magnetic fields and demonstrating 
power flares triggered likely by these fields. Observations of such 
flares reveal the
presence of quasi-periodic 
oscillations (QPOs) at certain frequencies; 
they are thought to be excited in the flares. 
QPOs carry 
potentially important information on magnetar structure, magnetic field, 
and mechanisms of magnetar activity. We calculate frequencies of 
torsional (magneto-elastic) oscillations of the magnetar
crust treating the 
magnetic field effects in the first order of perturbation theory. 
The theory predicts splitting of 
non-magnetic oscillation frequencies into Zeeman components.
 Zeeman splitting of torsional oscillation spectrum of magnetars
was suggested, clearly described and estimated by Shaisultanov and Eichler (2009)
but their work has not been given considerable attention. To extend it we suggest the
technique of calculating oscillation frequencies including Zeeman splitting at not
too strong magnetic fields for arbitrary magnetic
field configuration.
Zeeman  splitting enriches the oscillation spectrum and simplifies theoretical
interpretation of observations. We calculate several low-frequency oscillations
of magnetars with pure dipole magnetic field in the crust. The
results qualitatively agree with 
low-frequency QPOs detected in the hyperflare of SGR 1806--20, and
in the giant flare of SGR 1900+14.}


\begin{document}
	%%%%%%%%%%%%%%%%%%%%%%%%%%%%%%%%%%%%%%%%%%
	%\setcounter{section}{-1} %% Remove this when starting to work on the template.

%%%%%%%%%%%%%%%%%%%%%%%%%%%%%%%%%%%%%%%%%%%%%%%%%%%%
\section{Introduction}
\label{s:introduc}

It is well known that neutron stars are most compact stars
and contain superdense matter in their interiors (e.g. \cite{ST1983}). These stars have 
massive liquid cores and thin envelopes.
The core contains
matter of supranuclear density; its equation of state (EOS)
and other properties are still not certain and remain 
the fundamental problem of physics and astrophysics.
The envelope above the core 
is also important;
its thickness is $\sim 1$ km and
its mass is $\sim 0.01\,\msun$. The 
envelope consists mostly of  
electrons and atomic nuclei. In addition, at densities $\rho$
larger than the neutron drip density $\rho_{\rm drip} \approx 4.3 \times 10^{11}$ \gcc, there appear quasi-free neutrons. The atomic nuclei in the envelope form   
Coulomb crystals (e.g. \cite{HPY2007}) which 
melt near the very surface layers of the star. 
The solidified envelope is called the crust; one distinguishes
the outer ($\rho< \rho_{\rm drip}$) and the inner
($\rho > \rho_{\rm drip}$) crust. The density of matter at the
crust-core interface is 
$\rho_{\rm cc}\sim 1.4 \times 10^{14}$ \gcc.

This paper studies torsional oscillations of
the star due to elasticity
of crystalline lattice. In a non-magnetic star, such pure torsion 
oscillations are confined in the crust.  Foundation of the theory was laid by Hansen and Chioffi \cite{1980Hansen},
Schumaker and Thorne \cite{1983ST}, 
and McDermott et al. \cite{1988McDermott} in 1980s. Later the
theory was developed in numerous publications  (see, e.g., \cite{2007Samuel,2009Andersson,2012Sotani,2013Sotani,2013aSotani,2016Sotani,2017Sotani,2017aSotani,2018Sotani,2019Sotani} and references therein).

The theory attracted great attention after the discovery 
of quasi-periodic oscillations (QPOs) in spectra of soft-gamma repeaters (SGRs) (see \cite{2005Israel,2006Watts,2011Hambaryan,
2014Huppen,
2014Huppenkothen,2018Pumpe}). SGRs belong to
a class
of magnetars. They are neutron stars possessing very strong magnetic fields
$B \sim 10^{15}$ G (e.g. \cite{OlKas2014,2015Mereghetti,2017KasB}) 
and demonstrating
flaring activity that was most probably triggered by these
fields. The activity is accompanied by the processes of enormous energy
release in the form of hyperflares, supeflares and ordinary flares.
The QPOs have been discovered at fading stages of such flares.
Seismic activity of SGRs was predicted by
Duncan \cite{1998Duncan} in 1998.

The detected QPO frequencies range from a few tens
of Hz to several kHz. This is just the range 
typical for theoretical torsion frequencies of 
non-magnetic neutron stars. Magnetar QPOs are separated
into low-frequency ($\lesssim$ a few hundred Hz) and high-frequency ones. The discovery 
of magnetar QPOs gave hope to develop
reliable methods of exploring magnetar structure and evolution by comparing observations with
elaborated seismological models.  

The seismology of magnetars has been studied 
by many authors (e.g., \citet{2006Levin,2007Levin}, 
\citet{2006Glampedakis},
\citet{2007Sotani,2009CD},
\citet{2009Colaiuda,2011Colaiuda,2012Colaiuda},
\citet{2011vanHoven,2012vanHoven},
Gabler et al. \cite{2011Gabler,
2012Gabler,2013Gabler,2013Gabler1,2016Gabler,
2018Gabler},
\citet{2014Passamon,2016Link}).
They considered
different possibilities. The first idea was  
treating magnetar QPOs by ordinary
torsion oscillations of neutron star crust nearly
unaffected by magnetar fields (e.g., \cite{2007Sotani,2012Sotani,2013Sotani,2013aSotani,2016Sotani,2017Sotani,2018Sotani,2019Sotani}
). Other models were based on 
magneto-elastic oscillations of 
the crust, where the elasticity of crystalline lattice was 
combined with the elasticity of magnetic
fields lines; such oscillations can spread outside the crust in the 
form of Alfv\'en waves (e.g., \cite{2006Levin,2007Levin,2006Glampedakis,
2007Sotani,2008Lee,2009CD,
2009Colaiuda,2011Colaiuda,2012Colaiuda,
2011vanHoven,2012vanHoven,
2011Gabler,
2012Gabler,2013Gabler,2013Gabler1,2016Gabler,
2018Gabler,
2014Passamon,2016Link}). In some cases
the presence of the lattice
becomes fully unimportant; then the oscillations are of purely Alfv\'enic type and
can easily spread outside the crust.  

Let us emphasize the remarkable paper by \citet{2009SE} 
who predicted the effect of Zeeman splitting of torsional oscillations 
in magnetar magnetic fields. Although the effect is clear and the paper is
physically transparent, it has not been given much attention. It is our aim here to fill this gap, emphasize the
importance of the effect and extend the consideration of \citet{2009SE}
(although our technique will be somewhat different as we discuss later). 	
Here we focus on low-frequency magneto-elastic oscillations confined mainly to the crust, including Zeeman splitting. The paper is organized as follows. 

In Section \ref{s:approach}
we suggest to calculate magneto-elastic oscillation
frequencies using the first-order perturbation
theory with respect to the magnetic terms. 
The basic equations are presented in Section \ref{s:approach} neglecting relativistic effects (for simplicity). 
Pure torsional oscillations
are outlined in Section \ref{s:puretors}. The first-order
perturbation approach is described in Section \ref{s:iterate} for
any $\B$-field geometry. It demonstrates Zeeman splitting  \cite{2009SE} of
pure torsional oscillation frequencies into Zeeman
components, greatly enriching the oscillation spectrum. In Section \ref{s:axialsym} we use the perturbation
approach to the case of fundamental low-frequency torsional
oscillations 
in a poloidal and axially symmetric $\B$-field. In Section \ref{s:GR} we modify the equations by
including relativistic effects. In Section
\ref{s:Bdipole} we consider the pure dipole $\B$-field configuration in a neutron star crust. In Section \ref{s:sgrs} these results are used 
to sketch possible
interpretations of detected low-frequency QPOs in
afterglows of the hyperflare of SGR 1806--20  as well of the giant flare of
SGR 1900+14. Finally, in Section \ref{s:conclude} we summarize 
the results and compare our approach with those available in the
literature.

\section{The approach}
\label{s:approach}

Consider elastic oscillations of a neutron star crust 
containing a magnetic field {\bf B}. At {\bf B}=0 these
oscillations are pure torsional and occur due to elastic
properties of Coulomb crystals of atomic nuclei in the
crust. In a magnetic field, Alfv\'en waves, owing to elasticity
of magnetic field lines, modify elastic properties of matter
and lead to combined magneto-elastic oscillations where both
elasticities contribute. We will restrict ourselves
to the case of not too strong field {\bf B}, so that the magnetic
effects can be treated as perturbation. In this case we retain
the term `torsional oscillation,' for brevity.  

We start with a simplified 
problem in which we describe the star neglecting
relativistic effects; we
include them later. Otherwise, we adopt standard
assumptions for studying magneto-elastic oscillations based on linearization
procedure. All the quantities involved are divided into
unperturbed ones and small perturbations. The non-perturbed star is taken stationary and spherically symmetric; ordinary
spherical coordinates ($r$, $\theta$, and $\phi$) are 
most convenient 
for specifying position vectors $\bm{r}$. Unperturbed quantities, like
mass density $\rho$, depend only on $r$. The magnetic field {\bf B}
is assumed to be not too high and treated as perturbation. It is
split further as ${\bf B}=\B + {\bf B}_1$, where $\B$ is the
static (given) magnetic field of the star, while ${\bf B}_1$ is 
a smaller field induced by the oscillations. The field $\B$
leads to small stationary perturbations which do not interfere
with oscillatory perturbations in the adopted linear approximation.
Here we focus on oscillatory perturbations.

Let ${\bf v}(\vr,t)$ be a velocity
of a matter element, and ${\bf u}(\vr,t)$ be a shift of its position with
respect to equilibrium position $\vr$ in a non-perturbed star; 
$t$ is time. For an oscillation mode with  frequency
$\omega$ in the linear approximation one has
\begin{equation} 
{\bf v}(\vr,t)={\rm e}^{{\rm i}\omega t}\vv(\vr) ,~~~~
{\bf u}(\vr,t) ={\rm e}^{{\rm i}\omega t} \vu(\vr),~~~~ \bm{ v}={\rm i}\omega\,\bm{ u},
~~~{\bf B}_1(\vr,t)={\rm e}^{{\rm i}\omega t}{\B_1}(\vr),
\label{e:uv}
\end{equation}
where italic vectors $\vv$, $\vu$ and $\B_1$ 
are (generally complex and stationary) amplitudes of perturbations; dot denotes time derivative.

It is well known (e.g., \cite{2006Glampedakis}) that the torsional (magneto-elastic) oscillations can be treated as incompressible ($\nabla \cdot \vv=0$;
$\nabla \cdot \vu=0$). Then matter elements oscillate along spheres
with fixed radius $r$ and density $\rho(r)$, with $v_r=0$ and $u_r=0$. The density perturbations are
negligible.
The equation of motion for a matter element reduces than to the stationary equation
\begin{equation}
    \rho \omega^2 \vu= {\bf T}_\mu + {\bf T}_{B},
\label{e:newton}     
\end{equation}
where 
the force terms ${\bf T}_\mu$ and ${\bf T}_{ B}$ (minus elastic forces per unit volume) come from the
elasticity of crystalline lattice and magnetic field lines, respectively.
Elastic deformations of the lattice are determined 
by the shear modulus $\mu(r)$ taken in the approximation
of isotropic solid. 
Any $i$-th Cartesian component of ${\bf T}_{\mu}$ is given by
\begin{equation}
    {\bf T}_{\mu i}=- \frac{\partial \sigma_{ik}}{\partial x_k}, 
\label{e:Tsigma}        
\end{equation}
where $\sigma_{ik}= \mu \,(\partial u_i/\partial x_k +\partial u_k/\partial x_i )$ is the shear stress tensor.

The magnetic force term  ${\bf T}_{ B}$ is obtained
in the standard manner. From the Maxwell induction equation (without
dissipation) in our case one has  $\B_1={\rm curl}(\vu {\bf \times} \B)$ and 
\begin{equation}
    {\bf T}_{ B}= \frac{1}{4 \pi}\, \B {\bf \times }
    {\rm curl}\,[{\rm curl}(\vu {\bf \times} \B)].
\label{e:TB}    
\end{equation}
  
Eq.\ (\ref{e:newton}), supplemented with
(\ref{e:Tsigma}) and (\ref{e:TB}), allows one
to find oscillation eigenfrequencies $\omega$ and eigenvectors $\vu(\vr)$ which we discuss later in
more detail. It is important that the ${\bf T}_\mu$ and ${\bf T}_{\rm B}$ terms are linear in
the same shift vector $\vu$. This couples the lattice and $\B$-field elasticities.
The vector $\vu$ (with $u_r=0$) plays role of the wave 
function of the problem. 

Let us mention one feature  
of Eq.\ (\ref{e:newton}). Namely, let us multiply (\ref{e:newton}) by the complex 
conjugated shift vector $\vu^*$ and integrate over the star. In this way we obtain
the equality  
\begin{equation}
	\omega^2=\omega_\mu^2+\omega_B^2,
\label{e:freqdecompose} 	
\end{equation}
where
\begin{equation}
	\omega_\mu^2= 
	\frac{\int \dd V\, \vu^* {\bf T}_\mu}{\int \dd V \, \rho |u^2|}, \quad
	 \omega_B^2= 
	 \frac{\int \dd V\, \vu^* {\bf T}_B}{\int \dd V \, \rho |u^2|}.
\label{e:omega-sigma-B}	   
\end{equation}   

Therefore, the true oscillation frequency $\omega$ is formally expressed via two
auxiliary frequencies, $\omega_\mu$ and
$\omega_B$. In the absence of the magnetic field,
one has $\omega=\omega_\mu$, that refers to a pure torsional oscillation due to elasticity of the lattice. In the absence of the lattice, one gets $\omega=\omega_B$, that describes oscillations due to Alf\'ven waves. If the  lattice and the magnetic field are present at once, both 
auxiliary frequencies depend generally on crystal and magnetic elasticities. Both integrals in the
expression for $\omega_\mu^2$ contain
positive definite self-conjugated operators and bilinear combinations of $\vu$ and $\vu^*$ which guarantees
that $\omega_\mu^2 >0$. It may be not true 
for the integral in the
nominator of $\omega_B^2$; in principle, $\omega_B^2$ 
may be a complex number. 

The decomposition of frequencies $\omega$, similar to
(\ref{e:freqdecompose}), was studied earlier 
by Schumaker and Thorne \cite{1983ST} (also see, e.g.,
\cite{2020KY,2023Yak}) for pure torsional
oscillations at $\bm{B}=0$ with the aim to separate contributions of elastic shear forces and 
centrifugal forces in $\omega_\mu^2$. It is not a surprise that similar decomposition
appears in the case of magneto-elastic oscillations.

\section{Pure torsional oscillations}
\label{s:puretors}

This case has been studied in many
publications starting from
\cite{1980Hansen,1983ST,1988McDermott} 
and developed further (e.g., \cite{2007Samuel,2009Andersson,2012Sotani,2013Sotani,2013aSotani,2016Sotani,2017Sotani,2017aSotani,2018Sotani,2019Sotani} and references therein).

Here we present a brief summary of these results which we will need later. Oscillation modes are
characterized by three numbers. They are: i) multipolarity $\ell=$2, 3, \ldots; 
ii) azimuthal number $m$ (which is an integer varying
from $m=-\ell$ to $m=\ell$); and iii) the number
$n=$0,1, 2, \ldots of radial nodes of  wave functions $\vu(r,\theta,\phi)$.

The functions $\vu_{n \ell m}$ and
eigenfrequencies $\omega_{\mu n \ell}$ can be found from Eq.\
(\ref{e:newton}) with ${\bf T}_B$=0. 
In view of spherical symmetry of our non-perturbed
star, the quantisation axis $z$ can be arbitrary. Accordingly, the eigenfrequencies do not depend on $m$, meaning that any frequency is degenerate; it is one and the same for (2$ \ell$+1) different state vectors $\vu$ with
fixed $n$ and $\ell$ but different $m$. The consequences of this
fact will be important for 
our analysis.

In the given formulation, the solution for
$\vu$ can be presented as
\begin{eqnarray}
 u_\phi (r,\theta,\phi) & = &	r Y_{n \ell}(r) 
\exp({\rm i}m \phi)\, 
 \frac{\dd\,P_{\ell}^m}{\dd \theta},
\label{e:uphi} \\
 u_\theta (r,\theta,\phi) & = &	r Y_{n \ell}(r) \exp({\rm i}m \phi)\,
\frac{{\rm i} P_{\ell}^m}{\sin \theta},
\label{e:utheta}
\end{eqnarray}	
where $P_\ell^m(\cos \theta)$ is an associated Legendre polynomial,
and $Y_{n \ell}(r)$ is a convenient dimensionless 
radial wave function which satisfies the equation
\begin{equation}
	Y_{n \ell }''+\left( \frac{4}{r}+ \frac{\mu'}{\mu} \right)\,	Y_{n \ell }'+
	\left[ \frac{\rho}{\mu}\,\omega_{\mu n \ell }^2
	-\frac{(\ell+2)(\ell-1)}{r^2}   \right]\,
		Y_{ n \ell }=0 .
\label{e:Y}	
\end{equation}
Here the primes  mean derivatives with respect to $r$.
Since pure torsional oscillations are localized
in the crystalline matter ($r_1 \leq r \leq r_2$),
Eq.\ (\ref{e:Y}) is solved with the boundary conditions $Y'(r_1)=Y'(r_2)=0$
(of zero traction). The solution gives
the eigenfrequencies $\omega_{\mu n \ell}$. It is
Eq.\ (\ref{e:Y}) which contains
information on microphysics of neutron star matter. The angular dependence 
of the wave
vectors $\vu$ is standard (universal for all
stars).

\section{The simplest iterative solution}
\label{s:iterate}

Now we suggest the simplest iterative procedure 
to include the effects of stellar magnetic field $\B$. We treat the shift vector $\vu_{n \ell m}$, given by (\ref{e:uphi}) and (\ref{e:utheta}), as the zero-order
solution of (\ref{e:newton}), and we treat the
term containing ${\bf T}_B$, as perturbation. 
Let us restrict ourselves by  
the first-order perturbation. The problem is
similar to finding first-order correction to a non-perturbed degenerate
energy level of a quantum-mechanical system in case the perturbation removes
degeneracy and splits the energy into a  series of sublevels (e.g., \cite{LL76}). 

In this case, it is sufficient to use only
($2 \ell+1$) zero-order wave functions corresponding to
a fixed non-perturbed  
oscillation frequency $\omega_{n \ell m}$.
Such functions are often called the initial
zero-order wave functions.
The set of these functions is not unique: 
any linear superposition 
\begin{equation}
	\widetilde{\vu} = \sum_{m=-\ell}^{+\ell}
	\alpha_m \vu_{n \ell m}
\label{e:newstate}	 
\end{equation}
with constant transformation coefficients $\alpha_m$ also describes an oscillatory
state with the same $\omega_{n \ell m}$. 

The iteration procedure prescribes 
to calculate the perturbation matrix T$_{m m'}$
[of dimension $(2 \ell+1) \times
(2 \ell+1)$ ] 
on the basis of the initial wavefunctions. In our
case, such matrix elements are given by
\begin{equation}
	{\rm T}_{m' m}= \int_{\rm crust} \dd V \, \vu_{n \ell m'}^*\, {\bf T}_B(\vu_{n \ell m}).
	\label{e:Tmatrix}    
\end{equation} 
Since we deal with zero-order wave functions, which are localized
in the crust, the integration domain has to be restricted by the
crust alone. 

At the next stage one should use Eq.\ (\ref{e:newstate}) and introduce the basis 
of `true' zero-order wave functions
$\widetilde{\vu}_\nu$ (labelled by an index $\nu$) which diagonalize ${\rm T}_{m' m}$.
Let us denote the diagonal values by
$\widetilde{\rm T}_{\nu}$. Their number is $(2 \ell+1)$;
some of them can be equal if the perturbation removes 
degeneracy not completely. Then the
oscillation frequency for the `true' state $\nu$ is
\begin{equation}
	\omega_{n \ell \nu}^2=\omega_{\mu n \ell}^2
	+ \frac{\widetilde{T}_\nu}{\int_{\rm crust} \dd V \, \rho |u^2|}.
   \label{e:omegaB1}	
\end{equation}
This means the splitting of $\omega_{\mu n \ell}$ by the magnetic field that is traditionally treated as the Zeeman effect. This equation is in accord with the more general Eq.\ (\ref{e:omega-sigma-B}) in which $\omega_B^2$ 
has to be identified
with the last term in (\ref{e:omegaB1}).

This procedure of studying the magnetic splitting of torsional oscillation 
frequencies is computationally simple for any static magnetic
field $\B(\vr)$ configuration, being mostly reduced to calculating and 
diagonalizing the matrix (\ref{e:Tmatrix}). The  
procedure 
gives much richer spectrum of eigenfrequencies than the
theory of non-magnetic torsional oscillations. 
%To the best of our knowledge, 
%this procedure has
%not been used for investigating magneto-elastic oscillations of 
%neutron stars. 

The disadvantage of the procedure is evident. It is certainly
restricted by not too high magnetic fields: the magnetic splitting 
of oscillation frequencies has to be smaller than the zero-order
torsion oscillation frequencies $\omega_{\mu n \ell}$ themselves. 
The suggested procedure needs only the magnetic field $\B$ located 
in the crust; only zero-order wavefunctions are
involved.
Undoubtedly, sufficiently high $\B$ breaks down this approximation; then  penetration of Alfv\'en perturbations to the neutron star core and
possibly to the magnetosphere may become essential. At large $\B$ the theory should
be modified but the imprints of degeneracy problem need to be
studied anyway. Another disadvantage of 
the iterative scheme is that, although its first
step is feasible, higher-order iterations seem too complicated to be useful.  

\section{Fundamental torsional oscillations in poloidal
axially symmetric  magnetic fields}
\label{s:axialsym}

By way of illustration, we apply the formalism of Section \ref{s:iterate}
to the case of fundamental torsional oscillations in the presence of a poloidal 
axially symmetric magnetic field configuration. In this case
\begin{equation}
   B_r=B_r(r,\theta),\quad B_\theta=B_\theta(r,\theta),
   \quad B_\phi=0.
\label{e:axialsymmetry}   
\end{equation}
Here, $z$ is the magnetic axis.
Since ${\bf\nabla} \cdot \B=0$, the field components $B_r$ and
$B_\theta$ are related by
\begin{equation}
     \frac{1}{r^2} \, \frac{\partial}{\partial r}\,(r^2 B_r)+
     \frac{1}{r \, \sin \theta} \, \frac{\partial}{\partial \theta}\,
     (B_\theta\,\sin \theta)=0.
\label{e:divB=0}
\end{equation}

The main problem is to calculate the matrix elements T$_{m m'}$ from
Eq.\ (\ref{e:Tmatrix}) with the wavefunctions (\ref{e:utheta}) and 
(\ref{e:uphi}), and to diagonalize the matrix. Luckily, a careful inspection of 
the integral in (\ref{e:Tmatrix}) shows that for the magnetic field
(\ref{e:axialsymmetry})
the matrix T$_{m m'}$ is diagonal 
on the basis of the initial
zero-order wavefunctions because the integration over $\phi$ 
reduces to
\begin{equation}
    \int_0^{2 \pi} \dd \phi \, \exp({\rm i}m-{\rm i}m')\phi=2 \pi \,
    \delta_{m m'},
\label{e:deltamm}
\end{equation}
where $\delta_{m m'}$ is the Kronecker delta.
Accordingly, the initial zero-order functions are identical to the
true zero-order functions. Then the magnetically split eigenfrequencies   
can be labelled by $\nu=m$ and readily given by Eq.\ (\ref{e:omegaB1}) with
$\widetilde{\rm T}_\nu={\rm T}_{mm}$. This solution can be rewritten in
the form 
\begin{equation}
	\omega_{n \ell m}^2 = \omega_{\mu n \ell}^2 + \omega_{B n \ell m}^2,
	\quad \omega_{B n \ell m}^2=\frac{{\rm T}_{mm}}{\int_{\rm crust} \dd V \, \rho |u^2|}.
\label{e:omegaB2}
\end{equation}

In addition, due to the symmetry with respect to equatorial plane
($\theta=\pi/2$), one has T$_{mm}=$T$_{-m -m}$. Accordingly, the
splitting of Zeeman states with azimuthal numbers $m$ and $-m$,
is equal, so that the magnetic sublevels can be labelled by
$m=0$, 1,\dots $\ell$ (with $\ell+1$ different sublevels in total).
The sublevel $m=0$ is nondegenerate, while those with $m>0$ 
are still degenerate twice. The latter degeneracy can be further removed by a
more complicated $\B$-field geometry.

Moreover, we restrict ourselves by considering fundamental
torsional oscillations; they are those with no radial nodes. Then
we always have $n=0$, so that we drop this index, for simplicity.
Therefore, our zero-order oscillation frequencies $\omega_\mu =\omega_{\mu \ell}$
depend only on $\ell$. At low $\ell$ these frequencies are known to be small 
enough to explain lowest frequencies of QPOs observed in afterglow of magnetar
flares (see Sect. \ref{s:sgrs} for more details). For fundamental torsional oscillations,
with a very good accuracy, the radial wave function $Y$ in Eqs.\ (\ref{e:utheta}) and (\ref{e:uphi})  is independent of $r$: $Y_\ell(r)\approx Y_0$, 
see \cite{2020KY,2023Yak}. 
Since the numerators and denominators in Eq.\ (\ref{e:omega-sigma-B}) are proportional to $Y_0^2$, this quantity just drops
out of consideration. 

Then $\omega_{B \ell m}^2$ in Eq.\ (\ref{e:omegaB2}) can be rewritten as
\begin{equation}
	\omega_{B \ell m}^2=\frac{ \frac{1}{4 \pi} \int_{\rm crust} \dd V \, I_B}
	{ \int_{\rm crust} \dd V \, \rho r^2 
	\left(P'^2 + \frac {m^2}{\sin^2 \theta}\,P^2 \right) } ,
\label{e:omegaB3}	
\end{equation}
where $P=P_\ell^m (\theta)$, and
\begin{eqnarray}
I_B& = & -(B_\theta^2 -B_\theta \, B'_{\theta} \cot \theta
-B_r B_\theta \cot \theta -r B_r B_{\theta ,r}\ \cot \theta) \,
P'^2 
\nonumber \\
&&-(B_\theta \, B'_{\theta}+B_r B_\theta+ r B_r B_{\theta ,r})
P'	P'' - B_\theta^2 P' P''' 
\nonumber \\
&&  + \left[  B_\theta^2 P'^2 - (B_\theta B_{\theta}' + B_\theta^2 \cot \theta+ 2 B_r B_r'+ B_r B_\theta + r B_r B_{\theta, r}) P P'    \right.
\nonumber \\
&&
+\left. B_r (B_\theta'+ B_\theta \cot \theta
+r B_{\theta,r}'+ r B_{\theta, r} \cot \theta -B_r'' + B_r' \cot \theta)\,P^2
\right] \,\frac{m^2}{\sin^2 \theta }.
\label{e:IB}
\end{eqnarray}
Here, the primes denote derivatives with respect to $\theta$. The
subscript $r$ after comma denotes derivative with respect to
$r$. Eq.\ (\ref{e:IB}) is obtained from (\ref{e:Tmatrix}) by
substituting (\ref{e:utheta}) and$~~$ (\ref{e:uphi}), taking
$Y(r)=Y_0$, and integrating over $\phi$ using (\ref{e:deltamm}). We have
also used (\ref{e:divB=0}) to rearrange some terms.
The denominator of (\ref{e:omegaB3}) can be further simplified using the
well known (e.g. \cite{1966Arfken}) property of associated Legendre polynomials $P_{\ell}^m(\cos \theta)$:
\begin{equation}
 \int_0^\pi \sin \theta \, \dd \theta \left(P'^2 + \frac {m^2}{\sin^2 \theta}\,P^2 \right) \equiv 
 \Xi(\ell,m)= \frac{2 \ell (\ell+1) (\ell+m)!}{(2 \ell+1)(\ell-m)!}.
 \label{e:Xi}
\end{equation} 

Then we can rewrite (\ref{e:omegaB3}) as
\begin{equation}
\omega_{B \ell m}^2=\frac{ \frac{1}{4 \pi} \int_{r_1}^{r_2} \dd r \,r^2
	 \int_0^{\pi} \sin \theta \, \dd \theta \, I_B}
{ \Xi(\ell,m)\,
\int_{r_1}^{r_2} \dd r \, \rho r^4 } =
\frac{ \frac{1}{4 \pi} \int_{\rm crust} \dd V \, I_B}
{ \Xi(\ell,m)\,
	\int_{\rm crust} \dd V \, \rho r^2 }.
\label{e:omegaB4}	
\end{equation}
The integral in the denominator is exactly the same as for
pure torsional oscillations at $\B=0$.
Thus, the problem of calculating Zeeman splitting of zero-order oscillation
frequency reduces to the integration of known function $I_B$
over $r$ and $\theta$ in the nominator. This integral is determined  
by the magnetic field configuration  (\ref{e:axialsymmetry}) and can
be taken for any assumed poloidal axially symmetric $B_r$ and $B_\theta$. Otherwise the integration is insensitive to microphysics in the neutron star
crust. Although we have studied fundamental torsional modes ($n=0$),
the results are easily generalized to modes with radial nodes ($n>0$). 

According to the above expressions,
$\omega_{B \ell m}^2$ is quadratic 
in $B$. Then the true oscillation frequency has the
form $\omega(B)=\sqrt{\omega^2(0)+\alpha B^2}$,
where $B$ is a characteristic $\B$-field value, and $\alpha$ is some constant.
Since our consideration is strictly valid at $\omega^2(0) \gg
\alpha B^2$, one gets $\omega(B) \approx \omega(0)
+\tfrac{1}{2} \alpha B^2/\omega(0)$, meaning a small correction to $\omega(0)$ quadratic in $B$ in the guaranteed applicability range. However, taking into account the generic rule (\ref{e:freqdecompose}), we will often retain the square
root, which can formally exceed the applicability limit; at $\omega^2(0) \ll
\alpha B^2$ one would get then $\omega(B) \propto B$ although
the validity of this expression is unclear.

Similar square-root expression for $\omega(B)$ has been used in various calculations of magneto-elastic
oscillation frequencies (e.g. \cite{2007Sotani,2023Sotani}) which neglected Zeeman splitting (note that in some cases the dependence $\alpha B^2$ 
was declared to be
inaccurate at large $B$, e.g., \cite{2013Gabler}). The factor $\alpha$ has often been
used as a convenient parameter to fit the calculated $\omega(B)$. Our approach gives exact prescription how
to determine $\alpha$ including ($m \neq 0$) or disregarding ($m=0$) Zeeman splitting at not too high $B$.

Let us mention that in
the pioneering work  on Zeeman
effect in magnetars
\citet{2009SE}  used an elegant formalism of spherical vectors (e.g. \cite{1988Varshalovich})
to calculate (in our notations) T$_{mm}$.  
They assumed constant $\B$ directed along the $z$-axis and constant $\mu(r)/\rho(r)$ in the magnetar crust, and demonstrated the main features
of Zeeman splitting in magnetars. In case of more sophisticated magnetic field configurations, using the formalism of spherical vectors would be more complicated.

\section{Relativistic effects}
\label{s:GR}

So far we have neglected relativistic effects (to make our theoretical sketch  
physically transparent) but  
for neutron stars these effects are
quantitatively important.

The first effect is due to Special Theory of Relativity.
Since the neutron star matter is essentially relativistic, one 
should replace (e.g. \cite{1983ST}) the mass density $\rho$ in the equations of motion by
the inertial mass density $\rho+{\rm P}/c^2$, where P is the pressure. 

Other effects are those due to General Relativity
(GR). The spacetime 
in and around a neutron star is curved. For our problem,
it would be sufficient to use the relativistic Cowling
approximation and the standard metric for
a non-perturbed spherically symmetric star 
\begin{equation}
   \dd s^2=-\exp(2 \Phi(r)) \, \dd t^2 +\exp(2 \Lambda(r)) \, \dd r^2 
   + r^2 (\dd \theta^2 + \sin^2 \theta \, \dd \phi^2).
\label{e:metric}   
\end{equation} 
Here, $t$ is a Schwarzschild time (for a distant observer); $r$ is a radial coordinate that has meaning of 
circumferential radius; $\theta$ and $\phi$ are ordinary
spherical angles which are not affected by GR here; $\Phi(r)$ and $\Lambda(r)$
are the two metric functions to be determined from the
Tolman--Oppenheimer--Volkoff equations (e.g. \cite{1983ST,HPY2007}).

Let $M$ be the gravitational mass of the star and $R$ be its
circumferential radius. Outside the star ($r>R$) the metric 
(\ref{e:metric}) reduces to the Schwarzschild metric with
$\Phi(r)=-\Lambda(r)=\tfrac{1}{2} \ln(1 - r_{\rm g}/r)$. Here, 
$r_{\rm g}=2GM/c^2$ is the Schwarzschild radius of the star, $G$ is the gravitational constant, and $c$ is the speed of light.

The space-time in a relatively thin and low-massive neutron star crust can be substantially curved, but the curvature
is nearly constant there. Therefore,
the expressions for 
the oscillation frequencies, 
obtained in previous sections,
are expected to be rather accurate in a reference frame comoving the crust.

As demonstrated in \cite{2023Yak}, while 
calculating pure torsional
vibration frequencies detected by
a distant observer, it is
quite accurate to use the metric functions constant though the crust,
\begin{equation}
\Phi=-\Lambda=\tfrac{1}{2} \ln(1 - x_{\rm g*}),
\label{e:metricfun}
\end{equation}    
where $x_{\rm g*}=2GM_*/(c^2r_*)$, $r_*$ is any fiducial 
radial coordinate within
the crust, and $M_*$ is the gravitational mass within a sphere of 
radius $r_*$. 

One can easily show that this approximation, being applied to the redshifted frequency $\omega_{\mu \ell}$ 
of a pure torsional fundamental mode, reduces to
\begin{equation}
	\omega_{\mu \ell}^2=(1-x_{\rm g*}) 
    \frac{\int_{\rm crust} \dd V\, \mu }{\int_{\rm crust} \dd V \, (\rho + {\rm P}/c^2)\, r^2},
\label{e:freqtorsn=0}    
\end{equation}
where the volume element $\dd V=r^2\, \dd r \,
\sin \theta \, \dd \theta \, \dd \phi$. The factor $(1-x_{\rm g*})$ can be naturally interpreted as due to
the gravitational redshift. According to
\cite{2023Yak}, the maximum deviation of the approximate frequencies (\ref{e:freqtorsn=0}) from the
values of $\omega_{\mu \ell}$ calculated in full GR
does not exceed a few per cent for any choice of $r_*$ (although
choosing $r_*$ near the crust-core interface is slightly more accurate).
We do not pretend to be very accurate here and accept this accuracy.

Moreover, we introduce similar
relativistic corrections to the 
redshfited  quantity 
$\omega_{B \ell m}^2$. Then instead of 
(\ref{e:omegaB4}), we will use
\begin{equation}
    \omega_{B \ell m}^2=
    (1- x_{\rm g*})\, \frac{ \frac{1}{4 \pi} \int_{\rm crust} \dd V \, I_B}
   {\Xi(\ell,m)\,
	\int_{\rm crust} \dd V \, 
	(\rho+{\rm P}/c^2)\, r^2 } .
	\label{e:omegaB5}	
\end{equation}
In numerical calculations we take
$r_{*}$ at the crust-core interface.
In any case (\ref{e:omegaB5}) is an approximation, which seems to be 
sufficient for a semi-quantitative analysis,
although exact derivation of $\omega_{B \ell m}^2$ in full GR would be desirable.

\section{The dipole magnetic field}
\label{s:Bdipole}

Let us apply the above formalism to a
pure dipole magnetic field in
the neutron star crust,
\begin{equation}
	B_r = B_0 \left(\frac{R}{r}\right)^3
	\cos \theta, \quad
	B_\theta = \frac{B_0}{2}\, \left(\frac{R}{r}\right)^3
	\sin \theta,
	\label{e:Bdipole}
\end{equation}
where $B_0$ is the field at the magnetic pole on the surface in the local reference frame.

In this case Eq.\ (\ref{e:IB}) reduces to
\begin{eqnarray}
	I_B& = &- \frac{B_0^2}{4} \left(\frac{R}{r}\right)^6 \left[  
	P'^2 (1+ 3 \cot^2 \theta)
	-3 P'P'' \cot \theta + P' P''' \right.
	\nonumber \\
	&& \left. +
	\frac{m^2}{\sin^2 \theta}
	\left(-P'^2-10\, P P' \cot \theta
	+ 8P^2 \cot^2 \theta   \right)         \right].
	\label{e:IBdipole}	
\end{eqnarray}

Let us substitute this expression into
(\ref{e:omegaB5}). The integral over
$r$ in the nominator is taken analytically, and we obtain
\begin{equation}
	\omega_{B \ell m}^2
	= \frac{{B_0^2 r_2^3 }\, \LARGE[ \left( {r_2/r_1}  \right)^3 -1  
		\LARGE]}{12 \pi \int_{r_1}^{r_2} \dd r \, (\rho+{\rm P}/c^2)\, r^4}\, \zeta_{\ell m}, \quad \zeta_{\ell m}=
	\frac{1}{B_0^2 \, \Xi(\ell,m)}\,
	\int_0^\pi \sin \theta \, \dd \theta \,I_B(R,\theta). 
	\label{e:omegaBdipole}
\end{equation} 
Note that $r_1=R_{\rm cc}$ is the radius
at the crust-core interface, and $r_2$ 
is taken slightly lower than  $R$ for convenience of calculations;
shifting $r_2 \to R$ in the final expression
would not affect the result; see, e.g. \cite{2020KY}.

The Zeeman splitting is seen to be governed by
the integrals over $\theta$ in
$\zeta_{\ell m}$. We have calculated 
them for $\ell$=2, 3, 4, and 5.
%, and 6.
In all these cases, $\zeta_{\ell m}$ 
is virtually exactly described by
the expression
\begin{equation}
	\zeta_{\ell m}=c_{0}(\ell)+c_2(\ell)m^2,
	\label{e:fit}	
\end{equation}
where the coefficients $c_0(\ell)$ and $c_2(\ell)$ are
listed  in Table \ref{tab1}.

In this approximation $\omega_{B\ell m}^2$ is
proportional to $B_0^2$. In the range of strict validity 
of our approach
$\omega_{B \ell m}$ is also  quadratic in $B_0$, as already stated in
Section \ref{s:axialsym}. 

\begin{specialtable}[H] 
	\caption{Coefficients
		$c_0(\ell)$ and $c_2(\ell)$
		which describe $\zeta_{\ell m}$ in Eq. (\ref{e:fit}).\label{tab1}}
	%%% \tablesize{} %% You can specify the fontsize here, e.g., \tablesize{\footnotesize}. If commented out \small will be used.
	\begin{tabular}{ccccc} %c}
	\toprule
	$\ell$	& 2	& 3 &   4   &  5  \\ % & 6 \\
	\midrule
	$c_0$ & 0		    & 2.6667	&  6.5455      & 11.487  
	%& 17.454 
	\\
	$c_2$ & 0.66667		& 0.13333	&  --0.08182      & --0.252 
	%& --0.294 
	\\
	\bottomrule
\end{tabular}
\end{specialtable}

At $m=0$ the
presented expressions should be valid for describing previous results,
where Zeeman splitting was neglected. Then the lowest frequency
$\omega_{ \ell 0}$ with $\ell=2$ is not affected by the magnetic field at all
($c_0(2)=0$ in our approximation, which
seems true only for the dipole field),
while other frequencies with $\ell>2$ increase as $B_0^2$.  would be
instructive to check that $c_0(2)=0$ for the dipole magnetic field from
previous calculations of $\omega_{20}(B)$ but they are
controversial: for instance, compare fig. 4 in \cite{2007Sotani} with
figs. 1 and 2 in \cite{2008Lee}.

\begin{figure}[H]
	%\centering
	%\widefigure
	\includegraphics[width=0.35\textwidth]{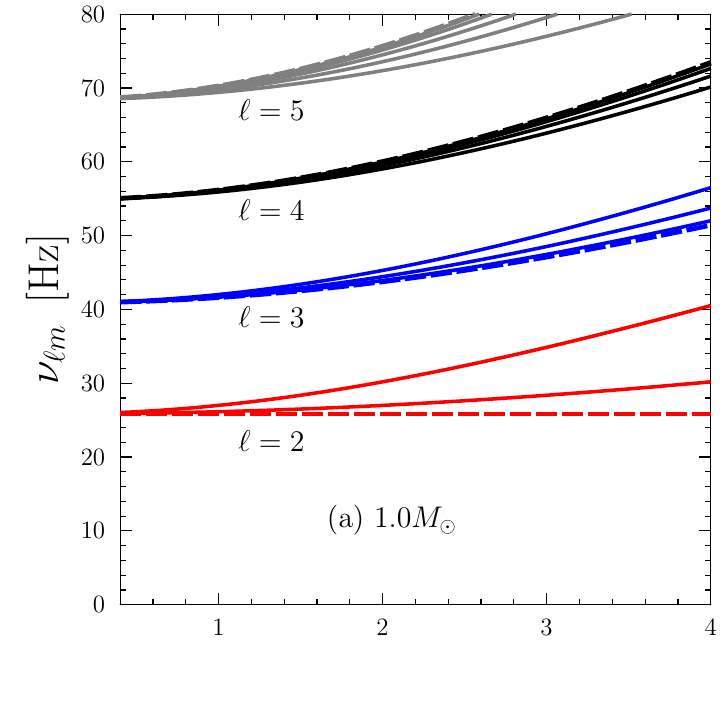}%
	%\hspace{8mm} 
	\includegraphics[width=0.35\textwidth]{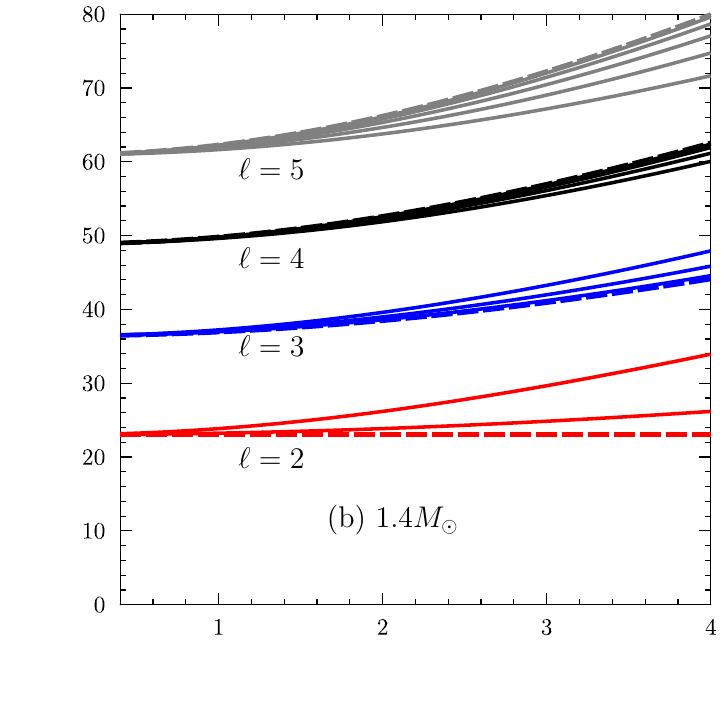}%
	\\	
	\includegraphics[width=0.35\textwidth]{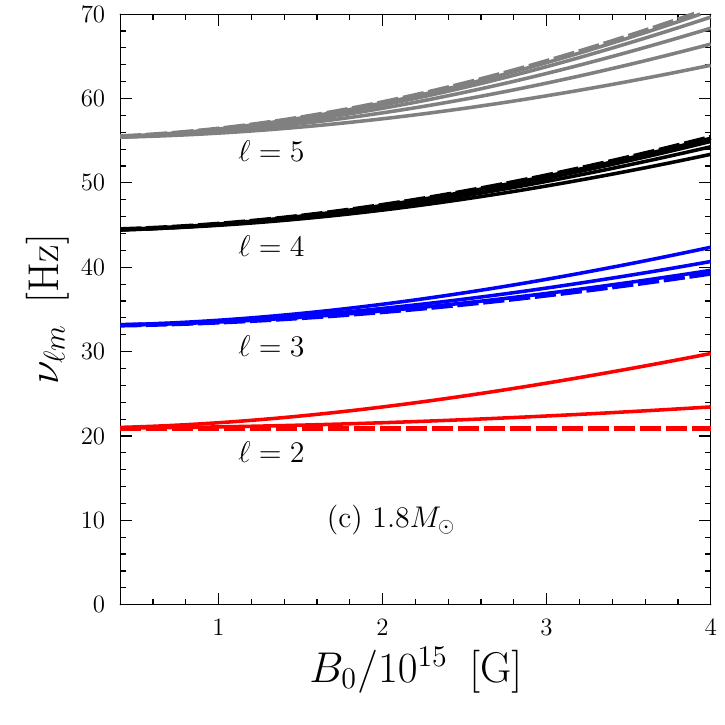}%
	%\hspace{8mm} 
	\includegraphics[width=0.35\textwidth]{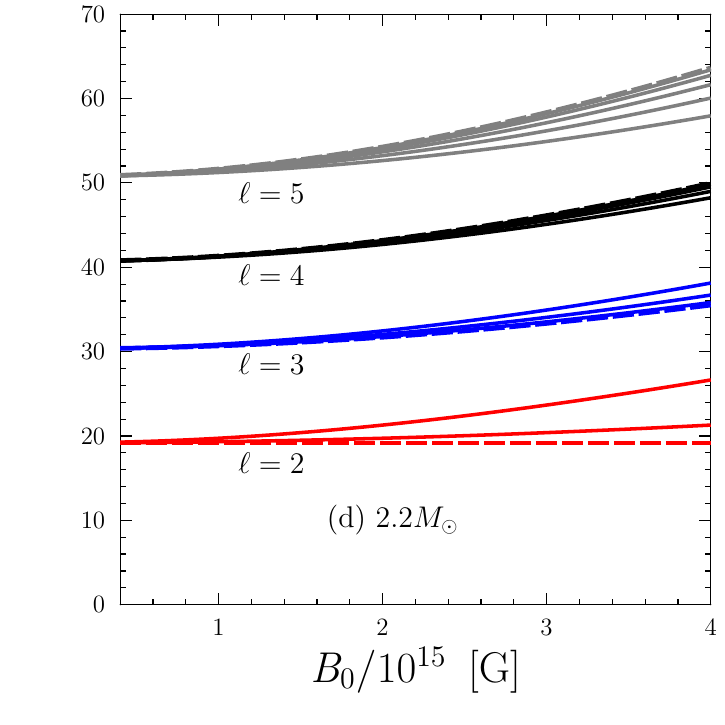}%	
	\caption{
		Fundamental torsional oscillation frequencies $\nu_{\ell m}=\omega_{\ell m}/(2 \pi)$
		at four lowest $\ell$ (from 2 to 5) 
		versus the dipole magnetic
		field $B_0$ at the magnetic pole. Calculations are performed
		for neutron star models with the BSk21 EOS
		at four masses; see panels (a)-(d).
		%(a) $1.0\,\Msun$, (b) $1.4\,\Msun$, (c) %$1.8\,\Msun$,
		%and (d) $2.2\,\Msun$. 
		One can observe Zeeman splitting
		of any frequency in a non-magnetic
		star into $(\ell+1)$ Zeeman components. The components with
		$m=0$ are plotted by dashed lines. With increasing $m$
		(at fixed $\ell$) the components either monotonically increase
		with respect to $\nu_{\ell 0}$ (for $\ell$=2 and 3)
		or decrease (for $\ell$=4 and 5).
	}
	\label{f:4M}
\end{figure}

%2    0       0.66667
%3    2.6667  0.13333 
%4    6.5455 -0.08182
%5    11.487 -0.252
%6    17.454 -0.294

The number of oscillation frequencies 
with $m=0$ is much lower than the total number of frequencies with different $m$. 
If we fix $\ell$ and vary $m$, the true oscillation frequency $\omega_{ \ell m}$
acquires the term behaving as $m^2 B_0^2$. For $\ell<4$ this term is positive,
but for $\ell \geq 4$ it becomes negative.

The presented expressions allow one to
calculate (evaluate) $\omega_{B \ell m}$ and, hence,  
total eigenfrequencies
$\omega_{ \ell m}$ of fundamental torsional 
(magneto-elastic) oscillations in the crust of 
any neutron star, whose model is given 
%assuming 
and the magnetic field is not too strong. If 
several oscillations  
frequencies are measured from one and
the same star, one can try to choose
a value of $B_0$ and a stellar model 
to explain a set of observed oscillations at once. 
This gives a method to constrain $B_0$
and the neutron star model.

For illustration, let us outline the main properties  of cyclic   
fundamental oscillation frequencies $\nu_{\ell m}=\omega_{\ell m}/(2 \pi)$ 
(expressed in Hz) at $\ell=$2,\ldots 5 and $m=$0,\ldots $\ell$ as functions of $B_0$. For 
this aim, we take 
neutron star models composed of matter with the BSk21 equation 
of state (EOS). This is a typical EOS; its basic 
properties as well as the properties of corresponding stellar models
are nicely described by analytic expressions by Potekhin et al.\
\cite{BSk2013}. The crust of these stars consists of spherical atomic
nuclei and electrons; the inner crust contains also quasi-free
neutrons, and there are quasi-free protons near the
crust-core interface ($\rho_{\rm cc}=1.34 \times 10^{14}$ \gcc).  
The cores of such stars are nucleonic, containing also electrons and muons. 
The crust and core are described by the same energy-density functional
of nuclear interaction. The maximum mass of the star of this type is
$M=2.27\,\Msun$. For instance, the stellar model with $M=1.4\, \Msun$
has radius $R=12.60$ km, and the radius of its crust-core interface is 
$R_{\rm cc}=11.55$ km. More information 
on neutron star models with the BSk21 EOS and
on pure torsional ($B=0$) oscillations of these
stars can be found in \cite{2023Yak}.

The behavior of the oscillation frequencies $\nu_{\ell m}$ with
increasing $B_0$ is shown in Figure \ref{f:4M}. The four panels
(a), (b), (c), and (d) correspond to the four neutron-star
masses 1.0 $\Msun$,
1.4 $\Msun$, 1.8 $\Msun$, and 2.2 $\Msun$, respectively. The field
$B_0 \leq 4 \times 10^{14}$ G is insufficiently strong to produce a noticeable Zeeman splitting with respect to $m$, and the oscillation
frequencies remain very close to those in non-magnetic stars.
At higher $B_0$ the Zeeman splitting becomes progressively more pronounced.
Each non-magnetic frequency with fixed $\ell$ splits
into $(\ell+1)$ Zeeman components. The components with lowest
$m=0$ are plotted by the dashed lines. These components have
been studied in the literature; the real spectrum is 
seen to be much more complicated. 
%The  frequency $\nu_{20}$
%is not affected by the magnetic field (in our approximation), while
%the frequencies $\nu_{\ell 0}$ with $\ell>2$ acquire the terms proportional to 
% $B_0^2$. 
 At $\ell=2$ and 3, the frequencies $\nu_{\ell 0}$ are
the lowest among respective Zeeman components ($\nu_{\ell m}$ increases with
$m$), while at higher $\ell$ they become the largest ($\nu_{\ell m}$ decreases with $m$). This  is reflected in inversion of sign of the coefficient $c_2(\ell)$ in Table \ref{tab1}.

According to Figure \ref{f:4M}, the dependence of 
$\nu_{\ell m}$ on $B_0$ for stars of different masses is similar. 
The frequencies $\nu_{\ell m}$ for more massive stars are 
smaller.
At sufficiently high $B_0$ our iterative solutions should become inaccurate
(because we extend the plots to very high $B$, see Sections \ref{s:iterate} and \ref{s:axialsym}).  For $\ell\leq 4$ this
seems to happen at $B_0 \gtrsim 4 \times 10^{15}$ G. With increasing
$\ell$ the Zeeman splitting gets wider, so that splittings for different $\ell$ 
start to overlap manifesting
possible cases of (avoided) crossings and mixtures of states
with different $\ell$. At still higher $\ell$ these effects are 
expected to occur at still lower $B_0$ resulting in a rich, densely spaced,  
and complicated spectrum of frequencies, a good project for 
further studies. 

\section{SGR 1806--20 and SGR 1900+14}
\label{s:sgrs} 

For illustrating the above results, let us sketch 
their use for possible interpretation of 
QPO oscillation frequencies observed in
X-ray afterglows of two flaring SGRs. Taken
the simplicity of our model we will not 
try to be accurate and seek the complete
sets of solutions, but present a few possible
cases chosen by eye. 

We analyze the low-frequency QPOs detected in the hyperflare of SGR 1806--20 (in 2004) and
giant flare of SGR 1900+14 (in 1998); e.g.,
\cite{2005Israel,2006Watts,2011Hambaryan,2014Huppen,2014Huppenkothen} and references therein. 

\begin{figure}[H]
	%\centering
	%\widefigure
	\includegraphics[width=0.35\textwidth]{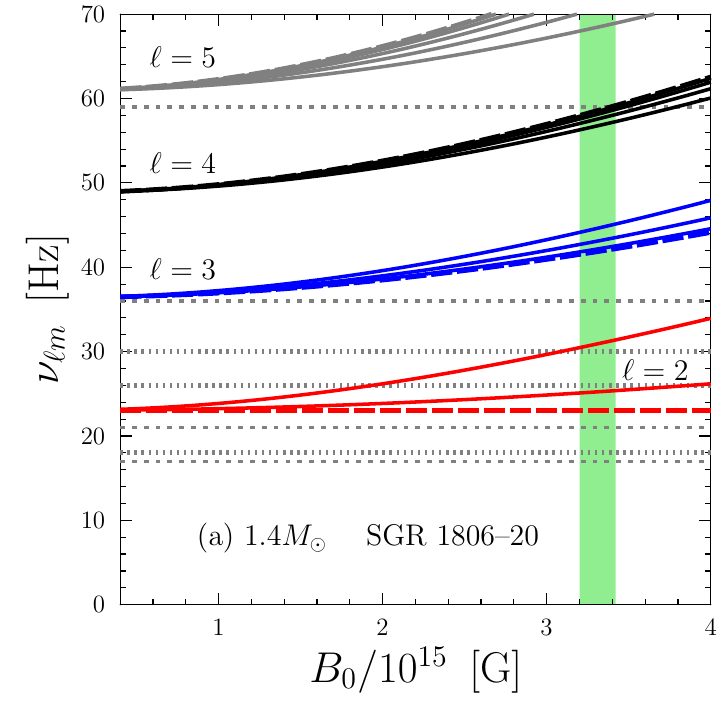}%
	%\hspace{8mm} 
	\includegraphics[width=0.35\textwidth]{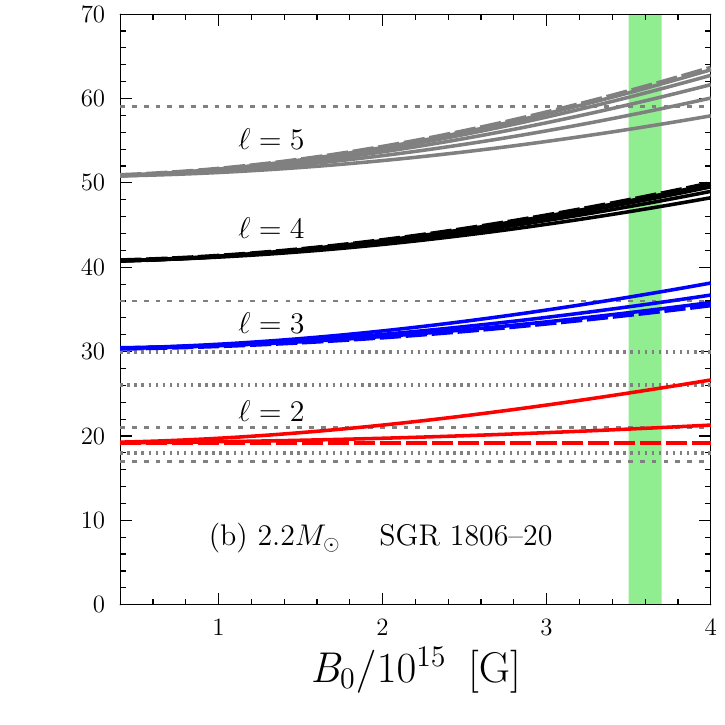}%
	%\\	
	% \includegraphics[width=0.35\textwidth]{figfreqn%at18.pdf}%
	%\hspace{8mm} 
	%\includegraphics[width=0.35\textwidth]{figfreqnat1806.pdf
	%}%	
	\caption{
		Calculated oscillation frequencies
		with $\ell \leq 5$ versus
		$B_0$ for two neutron star models,
		(a)  $M=1.4 \, \Msun$ (same as
		in Fig.\ref{f:4M}b) and 
		(b) $M=2.2 \, \Msun$ (same as
		in Fig.\ref{f:4M}d) compared with the
		low-frequency QPOs detected 
		in the afterglow of the hyperflare
		of SGR 1806--20. The detected frequencies
		are plotted by horizontal dotted
		lines. More reliable detections (18, 26 and 30 Hz) are plotted by denser dotted lines. 
		Less reliable ones (17, 21, 36 and 59 Hz) are displayed by more rarefied dotted lines.
	    Vertical shaded
		strips show possible ranges of $B_0$
		simultaneously consistent with some detections (see the text for details). 
	}
	\label{f:SGR1806}
\end{figure}

In particular, the low-frequency QPOs  in the spectrum of the hyperflare of SGR 1806--20 were detected at 18, 26, 30, 92 and 150 Hz (and
with less confidence at 17, 21, 36, 59 and 116 Hz).
Taking into account a restricted range of
frequencies calculated here (Fig.\ \ref{f:4M}), 
in Figure \ref{f:SGR1806} we analyze the possibility to simultaneously detect seven QPOs at
18, 26, 30, 17, 21, 36 and 59 Hz from one
and the same star with a dipole magnetic field
in the crust (leaving more complete
analysis for future studies). Naturally, the Zeeman splitting 
of oscillation frequencies greatly increases 
the probability of such interpretation.

Figure \ref{f:SGR1806}a shows the theoretical frequencies for the star of 
a typical mass 1.4\,$\msun$. By varying
$B_0$, we can choose such a $B_0$-interval that is consistent with more detections.
This interval, $B_0
\approx (3.2-3.4) \times 10^{15}$ G, is shown by a vertical bar. It allows
one to be qualitatively consistent with the three detected frequencies: 26, 30 and 59 Hz. Other
four frequencies are not explained in this way.
In particular, the frequencies $\lesssim$20 Hz remain unexplained. 

A more attractive solution can be obtained by taking 
a more massive star, where torsional oscillation
frequencies are smaller (e.g. \cite{2007Sotani,2023Yak}). This is shown
in Fig.\  \ref{f:SGR1806}b, where we take
$M=2.2~\msun$. With $B_0 \approx (3.5-3.7) \times 10^{15}$ G, we can now qualitatively explain the four detections at 18, 21, 26 and 59 Hz,
out of six. The detected QPO at 17 Hz is quite
close to QPOs at 18 and 21 Hz; it might be
produced by nonlinear interactions of QPOs at 18 and 21 Hz, especially if the 
magnetic field configuration is more complicated than the pure dipole. However, the QPO at 30 Hz
remains not explained. In any case, the
explanation in Fig.\  \ref{f:SGR1806}b seems more satisfactory than in Fig.\  \ref{f:SGR1806}a,
indicating that SGR 1806--20 may be a massive
star. The obtained field strength is
in line with current estimates of magnetar
fields (e.g., \cite{2017KasB}).   
 
\begin{figure}[H]
	\centering
	%\widefigure
	\includegraphics[width=0.35\textwidth]{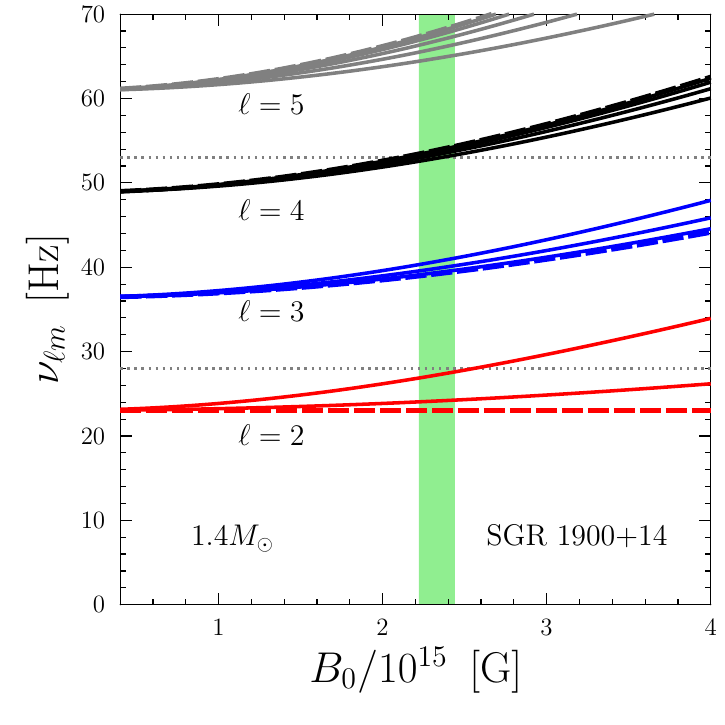}%
	\caption{
		Theoretical oscillation frequencies
		with $\ell \leq 5$ versus
		$B_0$ for a neutron star model
		with  $M=1.4 \, \Msun$ (same as
		in Fig.\ref{f:4M}b) compared with two low-frequency QPOs
		(28 and 53 Hz) detected 
		in the afterglow of the giant flare
		of SGR 1900+14  
		(the horizontal dotted
		lines). 
		Vertical shaded
		strip shows possible ranges of $B_0$
		consistent with both detected frequencies. 
	}
	\label{f:SGR1900}
\end{figure}

Figure \ref{f:SGR1900} compares our theory 
with the detection of low-frequency QPOs in
the giant flare of SGR 1900+14. The detected
frequencies were 28, 53, 84 and 155 Hz; our
{current results} can be used for explaining QPOs at 28 and 53 Hz. This
problem of interpreting two detected points
is much simpler than for SGR 1806--20. 
For instance, we take the $1.4~\msun$ star and
show one possible explanation, where
the field strength
$B_0 \approx (2.2-2.4) \times 10^{15}$ G 
is consistent with current
estimates of magnetar fields.

The results of this section 
can be treated as very preliminary. They just illustrate the importance
of Zeeman splitting of torsional oscillations 
of magnetars suggested by \citet{2009SE}.  

Let us stress once more that our approach is asymptotically accurate at not too strong 
$\B$-fields. However, our attempts (in this section) to interpret magnetar QPOs
indicate that real magnetar fields are just
those, at which our approach can be already broken. Finding accurate values
of breaking fields requires much effort. 

The approach is simple because it is valid in the regime in which
the elasticity of crystal matter is meant to be more important than
that due to magnetic field. Then the effects of bulk viscosity $\mu$ dominate
and the magneto-elastic oscillations have much in common with pure torsional 
oscillations at $B=0$. In particular, the oscillations are
mainly confined in the crystalline matter; in this way they depend on
the microphysics of the matter and the magnetic field
configuration in the crust. This allows one to
present solutions in nearly closed
form.

The difference of these solutions from those which were
extensively studied previously is that they deal with not axially symmetric
perturbations of matter velocities and magnetic field perturbations
(although the non-perturbed $\B$-field is axially symmetric). It  brings into
play with new degrees of freedom and
enriches the oscillation spectrum. It would 
be an interesting project to extend previous numerical simulations
(e.g. \cite{2012Gabler,2013Gabler}) to this case.

 With increasing $B$ the contribution of Alfv\'enic perturbations to
the oscillations will increase (and finally dominate); these perturbations will spread out of crystalline crust and the presented approach will break.
We fully appreciate many
previous investigations (e.g. \cite{2009CD}) of magnetar oscillations based on Alfv\'en wave propagation over the entire star (not confined in the crust). 
Such waves can produce sophisticated 
oscillations of the star which can be important for magnetar QPOs.  Currently this approach is mainly restricted
by axially symmetric deformations. It would be interesting 
to extend such studies to non-axially symmetric ones.

\section{Discussion and conclusions} 
\label{s:conclude}

We have studied fundamental torsional (magneto-elastic) oscillations of neutron star crust in magnetars. The solution at $\B=0$ is well known and corresponds to pure torsional oscillations of the stellar crust (Section
\ref{s:puretors}). The solution
for non-zero $\B$ has been extensively studied since the 
beginning of 2000s (e.g., \cite{2006Levin,2007Levin,2006Glampedakis,
	2007Sotani,2008Lee,2009CD,
	2009Colaiuda,2011Colaiuda,2012Colaiuda,
	2011vanHoven,2012vanHoven,
	2011Gabler,
	2012Gabler,2013Gabler,2013Gabler1,2016Gabler,
	2018Gabler,
	2014Passamon,2016Link}), mostly
numerically and for axially symmetric $\B$-field configurations assuming
axially symmetric velocity field of matter elements  and magnetic field perturbations.  
	
The exclusion was
made by \citet{2009SE} who considered not axially symmetric perturbations
and predicted the effect of Zeeman splitting of torsional oscillation frequencies
in magnetars. Unfortunately, their publication has not been
given considerable attention.

We have tried to extend their consideration basing on the same first-order perturbation
theory with respect to the magnetic terms in the linearized oscillation
equations. The equations are outlined in Section \ref{s:approach} neglecting relativistic effects. 
The properties of pure torsional oscillations
are summarized in Section \ref{s:puretors}. The standard first-order
perturbation approach is described in Section \ref{s:iterate} for
any $\B$-field geometry. The approach reproduces
 Zeeman splitting \cite{2009SE} of
pure torsional oscillation frequencies $\omega_{\ell n}$ into a bunch
of components, enriching thus theoretical spectrum of magneto-elastic
oscillations. We stress the well defined nature of the solutions, and 
their breakdown at very strong magnetic fields.

Section \ref{s:axialsym} applies the formalism of Section  \ref{s:iterate} to the case of fundamental oscillations (without nodes of radial
wave functions) in a poloidal axially symmetric $\B$-field. This case
is especially simple. Finding the oscillation frequencies reduces
to taking 2D integrals (\ref{e:omegaB3}) with well-defined integrands
(\ref{e:IB}). In Section \ref{s:GR} we modified the equations to
include relativistic effects. For illustration, Section
\ref{s:Bdipole} considers the simplest dipole field configuration in a neutron star crust. 
%For multipolarities $\ell\leq 5$ the magnetic
%correction (\ref{e:omegaBdipole}) to a pure torsional oscillation
%frequency is calculated exactly; see Eq.\ (\ref{e:fit}) and Table \ref{tab1}.  %This allows one to easily calculate magnetically 
%corrected fundamental torsional oscillation frequencies $\omega_{\ell m}$
%with $\ell \leq 5$ for neutron stars with different
%parameters; see Figure \ref{f:4M}.} 
In Section \ref{s:sgrs} our results are used 
for sketching possible
interpretations of detected low-frequency QPOs ($\leq 70$ Hz) in
afterglows of the hyperflare of SGR 1806--20 and of the giant flare of
SGR 1900+14. 
%For each SGR, we have tried to explain the detections using neutron
%star models of specific mass with certain values of the
%dipole $\B$-fields at magnetic poles. In case of SGR 1806--20 our
%preferred example refers to the star with $M=2.2 \, \msun$ 
%(the BSk21 EOS)
%and  $B_0 \approx (3.5-3.7) \times 10^{15}$ G
%(Fig. \ref{f:SGR1806}b). In case of SGR 1900+14, 
%one can assume $M=1.4 \, \msun$ and  $B_0 \approx (2.2-2.4) \times 10^{15}$ G; %see Fig.\ \ref{f:SGR1900}.

Let us emphasize that the Zeeman effect complicates the theory but greatly enriches the oscillation spectrum and simplifies in this way theoretical interpretation of detected QPOs. Needless to say that our restricted first-order iterative approach is far from being prefect and can be elaborated. First of all, one 
can consider other $\B$-field geometries (e.g. \cite{2008Aguilera}),
not pure dipole, but dipole-like fields, magnetic quadrupole, 
toriodal configurations, or mixtures of these, etc.  (see, e.g., \cite{2012Gabler,2013Gabler}). Preliminary estimates show
that for different $\B$-field geometries the Zeeman splitting can be 
qualitatively similar to that for the magnetic dipole 
but different in details. With the
loss of magnetic field symmetry, the splitting will be more complete,
removing residual degeneracy of oscillation frequencies. Additionally, one can replace approximate description of magneto-elastic oscillations in GR (Section \ref{s:GR})  by the exact description, in the spirit of
Ref.\ \cite{2007Sotani}. Moreover, one can extend the iterative 
approach to ordinary torsional oscillations, with finite amount of nodes of radial wave functions. Another direction of study could 
be to include delicate details of microphysics of
crustal matter like advanced shear modulus, 
the effects of superfluidity, possible nuclear pasta phases and
so on (as detailed, e.g., in \cite{2023Yak}).

%%%%%%%%%%%%%%%%%%%%%%%%%%%%%%%%%%%%%%%%%%
\vspace{6pt} 

%%%%%%%%%%%%%%%%%%%%%%%%%%%%%%%%%%%%%%%%%%
%% optional
%\supplementary{The following are available online at \linksupplementary{s1}, Figure S1: title, Table S1: title, Video S1: title.}

% Only for the journal Methods and Protocols:
% If you wish to submit a video article, please do so with any other supplementary material.
% \supplementary{The following are available at \linksupplementary{s1}, Figure S1: title, Table S1: title, Video S1: title. A supporting video article is available at doi: link.} 

%%%%%%%%%%%%%%%%%%%%%%%%%%%%%%%%%%%%%%%%%%
%DGY This shoulf be omitted: I am the only author
%\authorcontributions{\hl{Both authors contributed equally.}}%mdpi:please confirm if these parts should be added. For research articles with several authors, a short paragraph specifying their individual contributions must be provided. The following statements should be used ``Conceptualization, X.X. and Y.Y.; methodology, X.X.; software, X.X.; validation, X.X., Y.Y. and Z.Z.; formal analysis, X.X.; investigation, X.X.; resources, X.X.; data curation, X.X.; writing---original draft preparation, X.X.; writing---review and editing, X.X.; visualization, X.X.; supervision, X.X.; project administration, X.X.; funding acquisition, Y.Y. All authors have read and agreed to the published version of the manuscript.'', please turn to the  \href{http://img.mdpi.org/data/contributor-role-instruction.pdf}{CRediT taxonomy} for the term explanation. Authorship must be limited to those who have contributed substantially to the work~reported.

\funding{This research was supported by the
Russian Science Foundation (grant 19-12-00133 P)}

%DGY This should be excluded
%\institutionalreview{\hl{  }}%In this section, please add the Institutional Review Board Statement and approval number for studies involving humans or animals. Please note that the Editorial Office %might ask you for further information. Please add ``The study was conducted according to the guidelines of the Declaration of Helsinki, and approved by the Institutional Review Board (or %Ethics Committee) of NAME OF INSTITUTE (protocol code XXX and date of approval).'' OR ``Ethical review and approval were waived for this study, due to REASON (please provide a detailed %justification).'' OR ``Not applicable'' for studies not involving humans or animals. You might also choose to exclude this statement if the study did not involve humans or animals.

%DGY This should be excluded
%\informedconsent{\hl{  }}%Any research article describing a study involving humans should contain this statement. Please add ``Informed consent was obtained from all subjects involved in the %study.'' OR ``Patient consent was waived due to REASON (please provide a detailed justification).'' OR ``Not applicable'' for studies not involving humans. You might also choose to exclude %this statement if the study did not involve humans.	%Written informed consent for publication must be obtained from participating patients who can be identified (including by the patients themselves). Please state ``Written informed consent %has been obtained from the patient(s) to publish this paper'' if applicable.

\dataavailability{The data underlying this article will be shared on
	reasonable request to the corresponding author.} 

\acknowledgments{ I am indebted to the anonymous referee
for pointing out the publication by \citet{2009SE}. I am grateful to N. A. Zemlyakov and A. I. Chugunov for
useful discussion of advances in studying shear modulus of neutron star crust.}

%\conflictsofinterest{\hl{N/A }} %The authors declare no conflict of interest.

%% Optional
%\sampleavailability{Samples of the compounds ... are available from the authors.}

%%%%%%%%%%%%%%%%%%%%%%%%%%%%%%%%%%%%%%%%%%
%% Only for journal Encyclopedia
%\entrylink{The Link to this entry published on the encyclopedia platform.}

%%%%%%%%%%%%%%%%%%%%%%%%%%%%%%%%%%%%%%%%%%%
%%% Optional
%%\clearpage
\abbreviations{Abbreviations}{
The following abbreviations are used in this manuscript:\\
	
\noindent 
\begin{tabular}{@{}ll}
EOS & Equation of state \\	
GR &  General Relativity (theory) \\
SGR & Soft Gamma Repeater \\
QPO & Quasi-periodic oscillation
\end{tabular}}
%%%%%%%%%%%%%%%%%%%%%%%%%%%%%%%%%%%%%%%%%%%
\end{paracol}
%%%%%%%%%%%%%%%%%%%%%%%%%%%%%%%%%%%%%%%%%%%
%% To add notes in main text, please use \endnote{} and un-comment the codes below.
%\begin{adjustwidth}{-5.0cm}{0cm}
%\printendnotes[custom]
%\end{adjustwidth}
%%%%%%%%%%%%%%%%%%%%%%%%%%%%%%%%%%%%%%%%%%
\reftitle{References}

\newcommand{\araa}{Ann. Rev. Astron. Astrophys.}
\newcommand{\aap}{Astron. Astrophys.}
\newcommand{\aj}{Astron. J.}
\newcommand{\apjl}{Astrophys. J. Lett.}
\newcommand{\apj}{Astrophys. J.}
\newcommand{\apjs}{Astrophys. J. Suppl. Ser.}
\newcommand{\apss}{Astrophys. Space Sci.}
\newcommand{\mnras}{Mon. Not. R. Astron. Soc.}
\newcommand{\nat}{Nature}
\newcommand{\pasa}{Publ. Astron. Soc. Aust.}
\newcommand{\pasj}{Publ. Astron. Soc. Jpn.}
\newcommand{\pasp}{Publ. Astron. Soc. Pac.}
\newcommand{\prc}{Phys. Rev. C}
\newcommand{\pre}{Phys. Rev. E}
\newcommand{\prl}{Phys. Rev. Lett.}
\newcommand{\prd}{Phys. Rev. D}
\newcommand{\qjras}{Q. J. R. Astron. Soc.}
\newcommand{\sovast}{Sov. Astron.}
\newcommand{\ssr}{Space Sci. Rev.}

\externalbibliography{yes}

%\bibliography{YakBibList}

%\label{lastpage}

%\end{document}

\label{lastpage}

\end{document}